# Transverse emittance measurement in 2D and 4D performed on a Low Energy Beam Transport line: benchmarking and data analysis

F. Osswald,[a,*] T. Durand,[b] M. Heine,[a] J. Michaud,[c] F. Poirier,[d] JC. Thomas,[e] E. Traykov[a]

[a]*IPHC, CNRS/IN2P3, Université de Strasbourg, Strasbourg, France*
[b]*SUBATECH, CNRS/IN2P3, IMT Atlantique, Université de Nantes, Nantes, France*
[c]*LP2I, CNRS/IN2P3, Bordeaux, France*
[d]*ARRONAX, Saint-Herblain, France*
[e]*GANIL, CEA/DRF-CNRS/IN2P3, Caen, France*

*E-mail:* francis.osswald@iphc.cnrs.fr

ABSTRACT: 2D and 4D transverse phase-space of a low-energy ion-beam is measured with two of the most common emittance scanners. The article covers the description of the installation, the setup, the settings, the experiment and the benchmark of the two emittance meters. We compare the results from three series of measurements and present the advantages and drawbacks of the two systems. Coupling between phase-space planes, correlations and mitigation of deleterious effects are discussed. The influence of background noise and aberrations of trace-space figures on emittance measurements and RMS calculations is highlighted, especially for low density beams and halos. A new data analysis method using noise reduction, filtering, and reconstruction of the emittance figure is described. Finally, some basic concepts of phase-space theory and application to beam transport are recalled.

# Contents

## 1 Introduction

Charged particle accelerators show continuous advances in energy and luminosity and push the performance of beam instrumentation to unprecedented levels. The measurement of energy, current intensity, spatial and phase resolution, for the best known, must be at least as effective as the machine itself to offer a correct diagnosis. The development of beam instrumentation therefore presents a major scientific challenge and is essential during design, prototyping, commissioning, and operation of the accelerators. The beam instrumentation produces a figure of merit allowing for a good understanding of malfunctions, quantifying and reducing beam losses, and therefore damage to equipment and the impact on the environment through waste reduction of activated material.

    One of the most sophisticated diagnostics is the emittance meter. This experimental technique allows many applications as rms-emittance measurement, definition of optical aberrations, emittance filamentation, 2D tomography, evaluation of the charged particle density, transverse distributions of the position, and incidence angle of the charges. The measurement of emittance allows the complete definition of the beam in the transverse phase-space but also the evaluation of its homogeneity, its transverse profile, its position and its intensity. Emittance figure analysis provides useful information on the behavior of the beam during transport and enables the adaptation of the beam between the different sections of the accelerator in order to optimize the transmission. According to Liouville's theorem [1], the charged particle density must be constant in 4D or 6D phase-space i.e. under ideal conditions such as conservative forces, time-invariant and linear fields, no acceleration, without interaction between particles and the residual gas. Many references question the constant 2D emittance case and highlight the variety of problems encountered with accelerators, ion sources, storage rings, beam transport lines, beam handling equipment, *etc*.

    The concept of beam emittance was first introduced in the early 1950s at CERN in order to allow for good adaptation to the acceptance of the proton synchrotron and avoid losses [2]. Although electron beam optics has progressed since the late 1890s with the discovery of the deflection by electric and magnetic fields (unlike the light beam) and later in the 1920s with electron lenses and the microscope, the concept of emittance has been replaced by brightness in particular for application to electron sources, see for example [3]. Significant work has been done since the 1960s in the accelerators domain to understand emittance growth and the relationship between beam aberrations and entropy. The Maxwell-Boltzmann equation forms the basis of the kinetic theory of gases i.e. for neutral molecules

and thus defines the velocity distribution for a gas at a certain temperature. Nevertheless, in the case of charged particles subjected to an electric or a magnetic field a thermodynamic description of the transverse motions and the relation with emittance have been established and allow for a better understanding of the sources of the disturbances and aberrations observed [4]. The concept was later taken up to explain emittance growth and beam thermalization by increasing the entropy of the particle distribution, see for example [5]. A new application has been recently proposed to reconstruct complete 4D transverse phase-space distribution from multiple 2D projections using the principle of maximum entropy. The usual transverse 2D scans obtained with parallel slits are completed by 2D scans with perpendicular slits (exit slit rotated in respect of entrance one) to obtain the inter-plane coupling coefficients of the 4D beam matrix by reconstruction of the 2D projections [6]. This tomographic reconstruction technique of the 4D transverse phase-space distribution from 2D projections using the maximum entropy at SNS HEBT has been extended from earlier developments conducted at LANL with 1D beam profile measurements [7, 8].

A wide variety of techniques and instruments are used to measure the emittance of ion beams. Considering 2D and 4D transverse emittance measurement of LEBT lines, the most common systems are slit-slit, slit-grid, pepperpot, and quad scan. The first three are limited to a beam energy of a few MeV and more rarely to a few 10 MeV. The quad scan (or 3-gradient method) consists in measuring the size of the beam according to a quadrupole field strength. It makes it possible to evaluate the rms-emittance but in general it does not give access to the distributions. Diagnosis is therefore more limited than with other systems. The pepperpot technique provides a measurement of 4D emittance i.e. with non-zero inter-plane correlation moments if the beam exhibits some transverse coupling in the phase-space. Nevertheless, most devices show limited beam power and resolution compared to slit systems. The slit-slit systems, among them the so-called "Allison" devices exhibit a wide dynamic range of beam intensity from $\approx$ 1 fA to a few mA, up to 10 kW in average beam power, and spatial and angular resolution better than 1 µm and 1 mrad respectively.

This article aims to compare two of the most common emittance meters, *i.e.* the first one is based on the Allison principle, capable of scanning low energy ion beams and providing high resolution of the 2D transverse phase-space distributions [9]. The second one is a pepperpot system based on the KVI design [10]. The device is composed of a mask having linearly distributed holes in the vertical direction and able to scan the beam horizontally in order to deliver a 4D transverse phase-space distribution. In addition, the data was analyzed and filtered in order to extract the signal from the noise. A new method that reduces noise and reconstructs emittance figures is presented.

## 2 Description of the instruments

### 2.1 Allison type emittance meter

2D emittance-meters have been developed at the IPHC for low energy beam transport (LEBT) since the 1980s. One of the first instruments designed, constructed, and operated by the institute was based on a scanning slit-slit device, 5 mm wide copper strips collecting the charges, and a PDP-11 system (Digital Equipment Corp.) [11]. The instrument was dedicated to the characterization of ion production from different sources in the low-energy beam transport line of the local accelerators. In the 2000s, a new system based on the Allison principle was designed and built by an IPHC team as part of the SPIRAL 2 project [12]. It is used to characterize low energy ion-beams, features a scanning slit-slit system, and is based on the Allison's principle [9]. The ion-beam is sampled by a slit producing a beamlet which is analyzed according to its angle of incidence and its energy. The analysis is carried out by an electrostatic deflector composed of two parallel plates and two symmetrical slits at the entrance and the exit of the device, see Fig. 1.

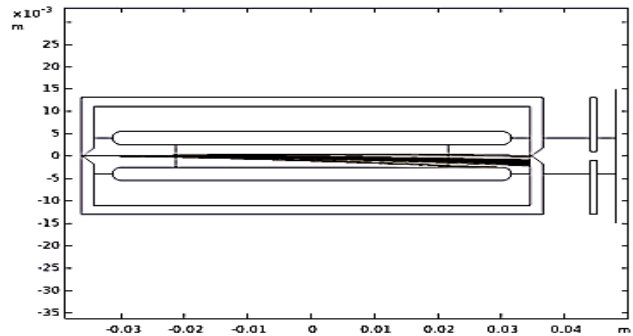

FIG. 1. Electrostatic analyzer is composed of a dipole and two symmetrical slits. The charge-to-mass ratio analyzer is completed with an electron repeller and a Faraday cup on the right side (beam enters on the left).

The four parts are symmetric and parallel. However, the ion-beam is unipolar, thus creating an asymmetry. It is attracted to one plate and repelled by the other, so beam is no longer symmetrical at the exit. Moreover, due to the Coulomb repulsion with high beam currents (space charge) and fringe field effect, the beam exhibits optical aberrations. Therefore, the standard relationship between the applied voltage, the ion energy, and the angle is valid with certain assumptions that are applied here. Phase-space transverse dimensions (*x or y* coordinates) are measured by saving the probe positions during the unidirectional beam scan. The intensity of the current is measured with a Faraday cup (FC) supplemented by an electron repeller located after the second slit at the exit of the device.

Between 2018 and 2020, IPHC developed with the collaboration of the FLNR and IN2P3 laboratories, a new prototype of 2D trace-space scanner based on the previous design of SPIRAL 2 with some improvements and sourcing to renew suppliers and reduce costs, see Fig. 2. The corresponding specifications are indicated in reference [13]. The collected charges are converted by a low-noise current-to-voltage preamplifier (transimpedance circuit). The low current intensity configuration is composed of a DLPCA-200 Femto module used to characterize the tail and the halo of the beams. The preamplifier has a gain of $10^3$-$10^{11}$ V/A and a DC-500 kHz bandwidth. The bandwidth depends on the intensity of the current. For low current intensities $\leq 1$ pA the BW is limited to 1 KHz. For higher current intensities with pulsed beams (10 mA total beam current, 0.1 mA on FC) the Femto DHPCA-100 module is used with a BW of 80 MHz. BW also depends on the capacitive charge (of the order of a few pF), and on co or tri-axial cable that imposes the RLC components. The signal is sampled at a rate of 100 kS/s then converted with a 16-bit ADC (National Instruments, NI 9215) with an input range of –10 to +10 V. A built-in FIFO memory allows for data buffering and continuous flow of measurements to the acquisition system. This allows measurements in time-slices over several pulses and therefore the observation of the dynamic evolution of the emittance. The samples are taken by a fast trigger on the ns scale allowing the reconstruction of the emittance over several pulses (1 to several 1000). The control system of the data acquisition and servomotor (brushless) is based on real time (RT) and FPGA programming. All other inputs and outputs (sensors, power supplies, etc.) are controlled with the National Instrument (NI) Compact RIO system (cRIO-9068, 9422, 9263, 9477, 9215), and a LabView GUI.

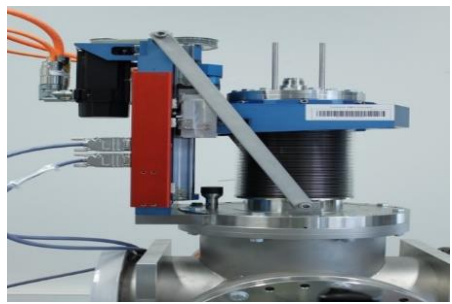

FIG. 2. Allison type emittance meter designed and constructed between 2018 and 2020 by IPHC and institutional partners, commissioned beginning 2021. On the left, the actuator composed of a driving motor, the guiding system and the feedings. On the right, the stainless-steel bellows and the flanges.

The system is now fully operational and has been following several experimental programs on different facilities since 2021. The main interest is to share a tool, the data and the experiences as part of the community at IN2P3 in France. It was used to characterize beams in the injection channel of the C70XP cyclotron at ARRONAX, Nantes [14]. Then it was installed at the exit of DESIR/ SPIRAL 2 High Resolution Spectrometer at LP2I Bordeaux, on the ARIBE beam line of the CIMAP/GANIL facility in Caen, and has been integrated during the Summer 2022 in a secondary beam line of the ALTO facility at IJCLab, Orsay, in order to characterize radioactive beams of new ion sources in the future. At the same time, a major investigation program has been carried out in order to understand the limits of the system and to propose improvements [13].

**2.2 Pepperpot type emittance meter**

The pepperpot-type 4D trace-space scanner is a Pantechnik system based on the KVI design [10]. The 1D sampling mask is a movable tantalum foil of 200 µm thickness with a row of linearly distributed holes (20 µm in diameter) and a pitch of 0.5 or 1 mm. This one-band design (as opposed to a 2D grid of holes) allows the measurement of higher beams intensities and smaller sampling steps in one axis with the guarantee of no data overlapping in comparison of the 2D mask design (better angular acceptance on one axis ≈80 mrad in comparison of 8 mrad for 1 mm step with the 2D mask). Hence, beam emittance can be measured in both transverse planes (XX' and YY') simultaneously with a better accuracy in the sampling plane (0.1 mm along scanning axis). On the other hand, the 1D mask design needs concatenation of the data and the acquisition time varies between 20 s and 2 min, depending on the desired resolution (step size).

Particles passing through the holes of the pepperpot mask are drifting over a fixed distance of 60 mm to a couple of polarized (≈1000 V) micro-channel plates (MCP) generating a cascade of electrons. Electrons are then accelerated toward a polarized (≈5000 V) phosphor screen (PS) to generate photons that can be visualized with a CCD camera. Photons are redirected with a 90° mirror so the camera can be placed perpendicularly to the beam axis, see Fig. 3. All the measuring system (mask + MCP + PS + mirror) can be remotely moved out of the beamline with a pressurized air system.

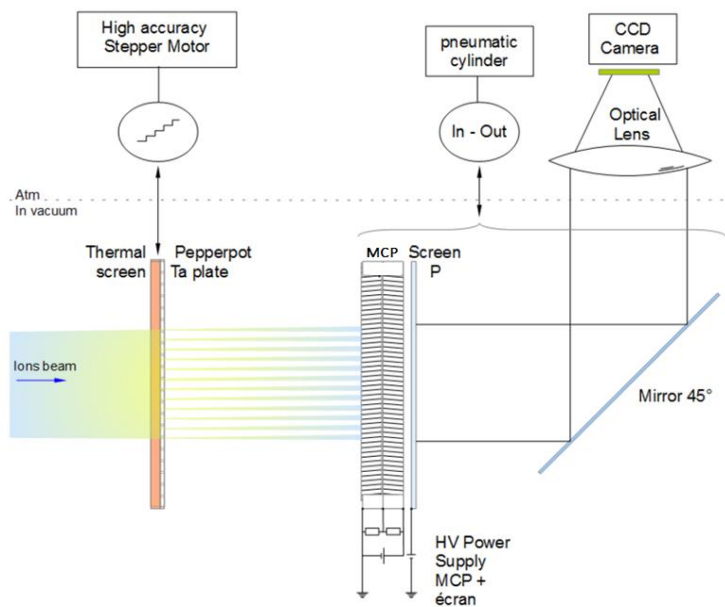

FIG. 3. Schematic diagram of the pepperpot emittance meter [15].

## 3 Experiment

Beams of 15 keV/q $^{40}$Ar$^{8+}$ from the ARIBE beam line at CIMAP/GANIL in Caen served to benchmark the emittance meters [16]. The beam intensity is typically 100 nA-1 µA DC. The beamline is equipped with an ECRIS [17], three dipole magnets (D), several quadrupoles (Q) and steering magnets (DC), see Fig. 4. The facility has a number of beam instruments such as beam profilers (PR), Faraday cups (FC), allowing for a control of the beam settings. The two emittance meters are installed on vacuum chambers near the plane of a waist and low dispersion, see Fig. 5. The beam optics simulation shows a particular beam setting obtained with smooth focusing, see Fig. 6. According to the references of the measurements that have been adopted, x is the horizontal plane, and y the vertical plane. The several optics elements and the instrumentation of the beam line enable the different beam setting configurations required by the tests. It is very challenging to make an absolute measurement of the emittance and more particularly of the momentum or incidence angle distribution of the charged particles. There are many parameters that influence the accuracy of the measurements and some settings require a calibration not just a setting. Beam settings are adapted to the requirements of each emittance meter, *i.e.* acceptance in position in the *x* and *y* planes, the angles of incidence, the scan range, and other related constraints as aperture and equipment of the beam line.

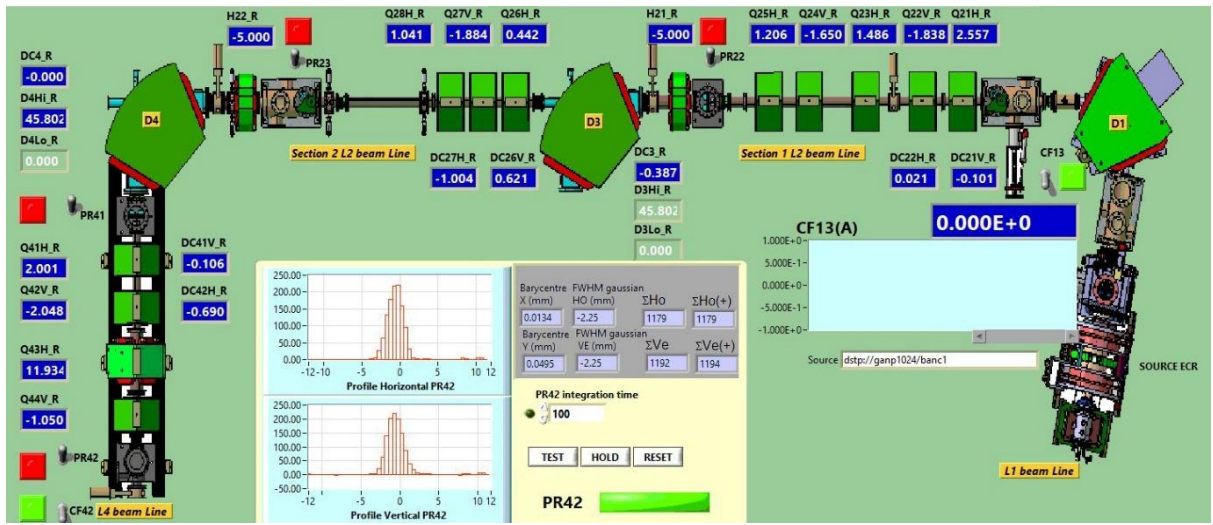

FIG. 4. Beam line layout of the ARIBE facility at CIMAP/GANIL. The blue windows show the optical elements settings and the larger one on the bottom shows the beam profile exiting the beam line (PR42). D1 and D4 dipoles are parts of an achromatic structure and D3 is switched off during experiment. ECRIS is positioned on the right and analyzed beam current is measured with CF13 Faraday cup (switched off).

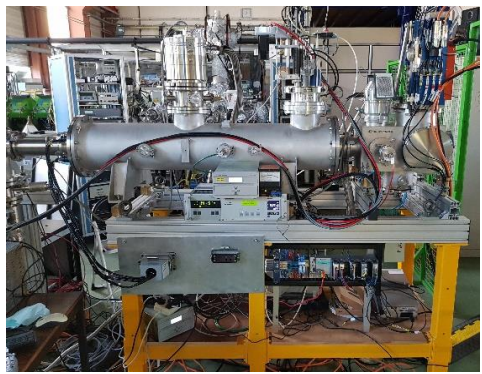

FIG. 5. Test stand with the two emittance meters installed close to each other but on separate vacuum chambers. The Allison type unit is installed on the small chamber on the right side, the pepperpot unit is installed on the large chamber before a turbomolecular pumping unit (TPU), both in vertical position (scan plane). The 120 keV $^{40}$Ar$^{8+}$ ion-beam is injected in the test stand from the left side.

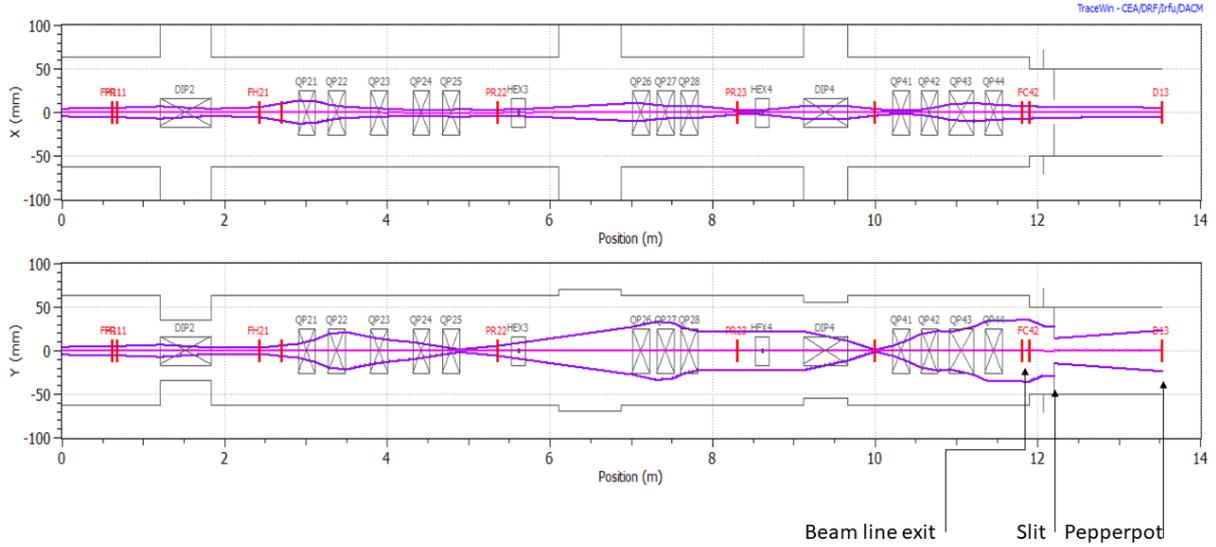

FIG. 6. Beam optics obtained with a TraceWin [18] model featuring the two main magnetic dipoles, the different quadrupoles labelled QP21 to QP44 and some higher order correction magnets (hexapoles). The last slit indicated in the figure allows to cut the beam and control transverse dimensions in addition to QP41-44. First emittance meter is the pepperpot unit installed near a waist. The transverse dimensions of the beam are adjusted to conform to the specifications of the two emittance meters (Allison type scanner located downstream not represented).

For the "Allison" emittance meter the scan range is 80 mm. For the pepperpot it is 15-30 mm because of the limited memory size and the constraint on the resolution, limiting the transverse dimensions of the beam in the corresponding plane. In the example of Fig. 6, the beam is convergent in *x* and shows a waist in the *y*-plane in the plane at the position of the pepperpot emittance meter. The beam diverges in the x and *y* planes at the position of the "Allison" emittance meter (not represented). Multiple beam settings were previously studied with TraceWin simulations, tuning magnets (Q41-44 in Fig. 4) and the last slit (Fig. 6). The beam line offers many theoretical possibilities to obtain the desired optics. For example, by defocusing the beam with Q44 instead of Q41, the vertical beam envelope of the beam is reduced at the position of the emittance meters and it shows a much greater divergence downstream. In this way, the beam is fitted within the range of both the emittance meters.

The two emittance meters are mounted on the same beam line section, not in the same plane, in a vertical position allowing a 2D phase-space distribution measurement in $yy'$ plane for the Allison emittance meter and a 4D phase-space distribution measurement with maximum resolution in the vertical plane for the pepperpot one. It is important to verify its rest position and remove the protection shield in order to guaranty the perfect transparency during the emittance measurements of the Allison type unit positioned downstream. The two emittance meters have the same references and are aligned with the same tools on the same reference orbit (beam axis). The alignment of the slit that samples the beam is of paramount importance. The greatest attention has been paid to the alignment of the two different measuring heads. The procedure for aligning the emittance meters is based on a laser beam and several fiduciaries fixed on the flanges which serve as a reference. All constituent components are aligned during assembly according to specifications and tolerances. The slit's reference position is fixed on the longitudinal axis with the laser during a fine scanning step and finally the complete scan over the full range is verified on the control panel (0.1 mm slit aperture). After installing the beamline equipment and pumping the vacuum to $10^{-6}$ mbar, the background signal is stored without beam to be subsequently subtracted from the emittance images.

The different settings of the pepperpot emittance meter are adjusted during the experiment in agreement with standard procedures [10, 15] and according to beam size, intensity and needs. A preliminary calibration sets the relation between the number of pixel and the transverse dimensions, typically 20 px/mm. The tuning is an important step of the experimental process as the final (absolute) emittance values will depend on a few settings:

i. MCPs polarization voltage has to be tuned according to the beam intensity, typically 1 kV.
ii. The phosphor screen offers the best signal/noise ratio for maximal polarization voltage, typically 5 kV.
iii. Non-linear effects due to MCP saturation and time resolution limitation due to PS decay time have been toned down with the best possible settings.
iv. A non-optimized focal point tuning of the CCD camera gives overestimated emittance values.
v. Due to system internal limitations, the beam must be focalized and centered on the pepperpot such that its envelope is entirely contained in the mask. The acquisition is limited to 51 data points in the scanning axis (here in the vertical plane).

During the installation of the pepperpot emittance meter, a calibration with the three-profile measurement method was carried out. This technique requires three sets of slits to determine the three unknowns $\varepsilon, \alpha, \beta$ and consists of geometrically constraining the emittance figure so that it forms a hexagon of known area. If the slits are respectively open with an aperture $2x_0, x_0, 2x_0$ with a beam waist on the central slits and separated by a distance L, the emittance can be estimated with following equation, see (1):

$$\epsilon = \frac{\sqrt{3} \cdot (x_0)^2}{L} \pi \text{ mm.mrad} \quad (1)$$

with $L$ = 900 mm and $x_0$ = 2 mm, the measured emittance of a reference beam according to equation (1) is 7.69 $\pi$ mm.mrad. Fig. 7 shows the emittance measured with the pepperpot emittance meter. Rms-emittance is 7.19 $\pi$ mm.mrad that indicates a difference of 6.5 % with the reference value.

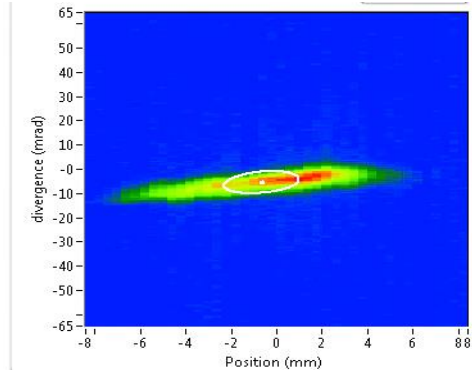

FIG. 7. 2D emittance figure obtained in the vertical plane *(yy')* during the installation of the emittance meter for the purpose of evaluating the accuracy of the measurement. Rms-emittance is 7.19 $\pi$ mm.mrad obtained with a well-calibrated beam.

## 4 Benchmarking

The IPHC's emittance meter of Allison type is benchmarked with the pepperpot system of Pantechnik to complement previous investigations [19]. Measurement of the transverse emittance is carried out in the vertical plane with different beam cuts allowing a variation of the emittance (plane defined as *yy'* for the Allison type emittance and *xx'* for the pepperpot system). The standard $^{40}Ar^{8+}$ DC beam at total kinetic energy of 120 keV and current intensity of 1 µA defined above is produced with an elliptic shape, a marginal emittance of $\approx$ 60 $\pi$ mm.mrad, and without optical aberrations. Beam dynamics simulations of backtracking are requested to compare emittances in the same vertical plane. The distributions in 2D and 4D phase-space are derived from measurements of the "Allison" and pepperpot emittance-meters. The simulations are carried out with the multiparticle code TraceWin. Simulated phase-space figures do not show distortions because there are no magnet and nonlinear effects in the drift space between emittance meters.

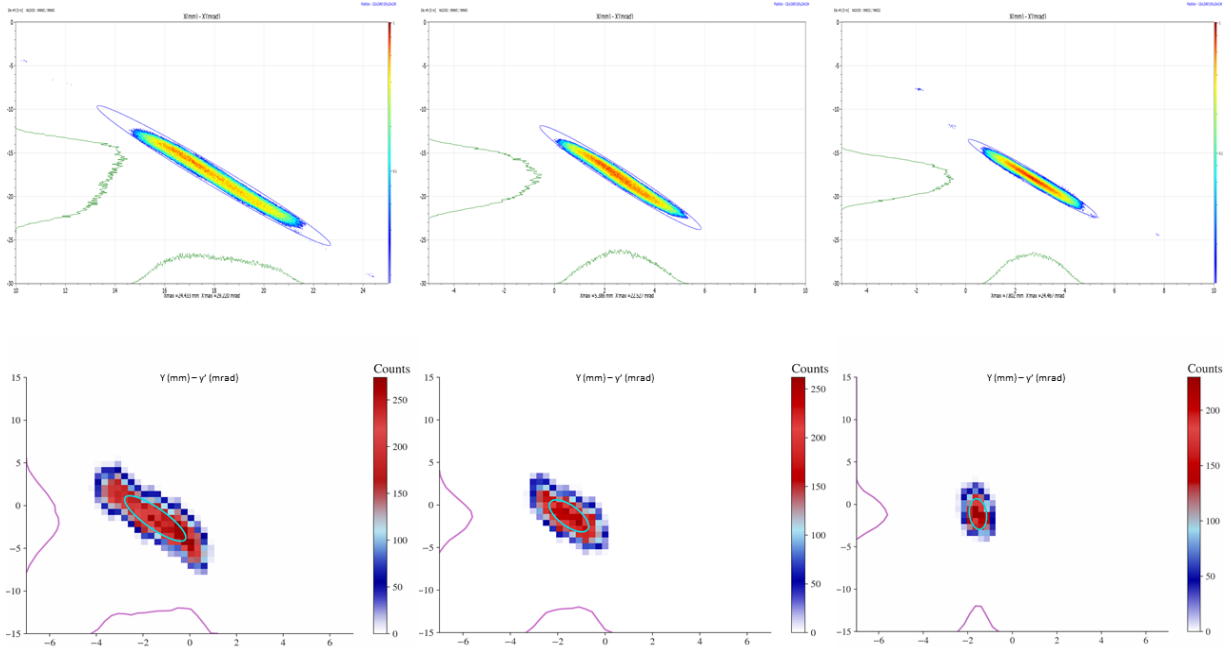

FIG. 8. Three series of emittance measurements carried out with the two instruments and three different beam settings. Each plot shows the 2D emittance in the vertical plane. The three plots on the top are obtained with the Allison type emittance meter and the following three plots on the bottom are obtained with the pepperpot emittance meter. The same threshold of 50 % was applied on all the data in order to eliminate the background noise. The data of the Allison emittance meter are displayed with TraceWin.

As shown in Fig. 8, results are contrasted. Disregarding the difference in resolution between the two systems, emittance figure shape, angle of inclination of the ellipse and 1D distributions are similar, except for the last run where emittance figures are different. The difference in the last series has been analyzed in detail and cannot be explained with a difference of focusing, a beam crossing the longitudinal axis, and other differences in the settings of the emittance meters. The slit opening allowing the beam cut-off and the emittance reduction is fixed at 7.5, 5 and 2.5 mm, see Fig. 6. The beam current intensity is reduced by a factor of 4 between the largest and smallest opening. The misalignment of the beam, visible on the figures, could not be reduced during the experiment. The misalignment is confirmed by the imbalance of the current collected on the two sides of the isolated slit. It should be noted that measurements are performed during the same day, during a period of one hour, and with equivalent delays for both the systems (≈30-60 min.). Thus, step numbers for the scans in position for the pepperpot, in position and voltage for the Allison type are equivalent.

Some differences are noticed between the RMS values calculated from the 2D distributions, see Table 1. The errors are due to the background noise (BG) generated mainly by secondary electron emission and the proximity between the emittance meters, and a beam dump at the end of the beam line. The maximum difference of 45% is obtained with a conventional background filtering which reflects its influence on the rms-emittance measurements as detailed in the following. It illustrates the difficulty of performing an absolute emittance measurement and comparing it with different systems.

Table 1. Comparison of rms-emittance values of the vertical trace-space distribution and Twiss parameters for the three series of measurement performed with the Allison type emittance meter (IPHC) and the pepperpot system (Pantechnik). The reduction of the RMS value is due to the slit opening variation and emittance cut-off (see Fig. 6).

|  | Series 1 | | Series 2 | | Series 3 | |
| --- | --- | --- | --- | --- | --- | --- |
|  | Allison | Pepperpot | Allison | Pepperpot | Allison | Pepperpot |
| $\varepsilon_{yy'}$ RMS ($\pi$ mm.mrad) | 0.72 | 1.32 | 0.44 | 0.80 | 0.28 | 0.33 |
| $\alpha$ | 5.84 | NA | 4.72 | NA | 4.72 | NA |
| $\beta$ (mm/$\pi$ mrad) | 3.47 | NA | 2.60 | NA | 2.81 | NA |
| RMS difference (%) | 45 | | 45 | | 15 | |
| Slit aperture (mm) | 7.5 | | 5 | | 2.5 | |

# 5 Data analysis

When considering LEBT, beam emittance measurements pay "tribute" to a large number of spurious signals with negative values, beam contaminant, electromagnetic (EM) interference, and other sources of disturbance. Several studies have revealed the importance of the background noise on the emittance measurement [20-22]. The background depends on the presence, specifications and working conditions of the surrounding equipment as power supplies, vacuum pumps, magnets, RF devices and actuator (ON/OFF, power, proximity, shielding). It also depends on the settings of the electronics (I/U convertor), data acquisition (number of samples, calculation of average values, etc.), and environment: vacuum level, secondary beam emission, and temperature.

A beam of non-elliptical shape, without a Gaussian distribution, featuring several species, the charge states and Coulomb interactions can lead to space charge effect and emittance growth. Some of the spurious signals, that can either be randomly distributed or cluster in the phase-space, overlap the beam of interest, particularly in the tail and halo region, making it difficult to distinguish between signal and noise and thus identify the background. The usual technique of background elimination by subtracting an image taken without beam from the data is not sufficient. When it comes to measuring a current intensity of the order of pA or less, the characterization of the tail and the halo of the beam is a real challenge. Simple techniques applying a threshold reduce the background but also remove relevant data. A program for the study of spurious signals and reconstruction of emittance pictures has been carried out in order to improve the identification of the background, and therefore to increase the measurement accuracy. Conventional data analysis such as data filtering, background subtraction, "thresholding" data, 2D Gaussian and polynomial fit, image reconstruction, pattern recognition and statistics tools have been used to define a new method applied to emittance measurements. In the following, two different signal identification and image reconstruction techniques are described.

## 5.1 Data analysis for the Allison type emittance meter

A specific development was carried out for the data analysis of the Allison type emittance meter in order to improve the quality of the emittance figures and of the RMS calculations. It aims to separate the signal from the background noise and to characterize the desired beam where standard parameters such as a 2D Gaussian fit of an embedded signal with adjusted 2D noise are utilized. Discrimination between signal and noise and image reconstruction are some techniques used to improve emittance measurements and RMS calculations. During a scan in the vertical trace-space over a range of 0-10 mm and 0-30 mrad, with 0.2 mm and 0.2 mrad steps, the acquisition generates 25 x 200 data points. Most of them have the same polarity but only a few of them concern the signal of interest.

Initial fits are used to identify the region of interest (ROI) of the signal and the background. The fit parameters are mitigated into a global fit including both the 2D Gaussian and a polynomial shape. The different steps of the new method are described in the algorithm in Fig. 9.

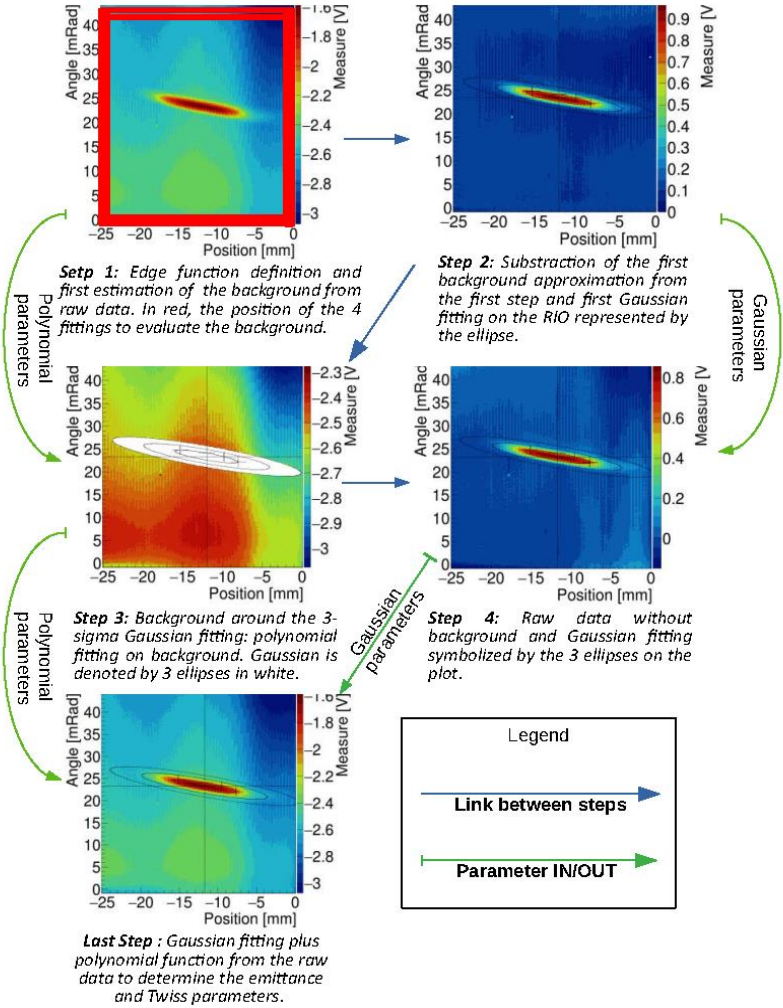

FIG. 9. Block diagram of the algorithm used for the new data analysis method applied to the Allison type emittance meter, with extraction of signal from background noise.

The Gaussian and polynomial fits are optimized iteratively. First, the background is fitted separately with higher order polynomial equations along the four sides of the 2D transverse phase-space (where no signal is expected), as highlighted in step 1 in Fig. 9. The equations are linearly combined to generate a defined surface for each data point (pixel). This step leads to the definition of the uncorrelated part of the background. The resulting area is subtracted from the raw data to reveal the signal, as shown in step 2. Next, a 2D Gaussian fit is applied to the cleaned signal in order to identify the ROI [23]. These parameters are used as inputs for the subsequent Gaussian fit resulting in the reliable convergence of the method, see step 3. It aims to provide a more precise definition of the background with a polynomial fit of the data outside the ROI. The data corresponding to BG are obtained by excluding the RIO of the signal (see white area) with a predetermined number of the RMS value (arbitrary number of Sigma are chosen, usually three or four Sigma). The result of the polynomial fit is subtracted from the raw data followed by a second Gaussian fit on the ROI with the previous Gaussian fit parameters as inputs during step 4. The last fit takes as input the previous result of the 2D Gaussian fit plus the polynomial function. The result of the final 2D Gaussian fit is used to calculate the emittance with the sigma of the 2D Gaussian fit [24] and the Twiss parameters with the emittance ellipse description [25]. This technique is designed to provide a reliable description of the background, thus to separate the signal from the background and to obtain a smoothing of the statistical fluctuations in the measurement.

The uncertainty of the angle, position and voltage amplitude measurements, respectively, are due to the Trek TZD700A high-voltage generator (DC voltage gain accuracy $\leq 0.1\%$ with a gain of 200 V/V, and output noise $\leq 75$ mV rms with a 1 nF load), the width of the emittance meter slits (aperture of 0.1 mm), and the NI acquisition card error (0.02% gain error with16-bits 100 kS/s analog input channels) are included in each fit.

In order to evaluate the limits of the new method, the same procedure is applied to other emittance measurements as around 10% of the beams don't resemble a 2D Gaussian shape. Limits can be identified by the error calculated with Pearson's chi-square of the fit. An example of critical beam shape is depicted in Fig. 10 (emittance figure) and Fig. 11 (fit residue). This can be interpreted by a large variation between the raw data and the fitting function, as shown in Fig. 11. The fluctuations are due to the Gaussian function used in the fitting equation which does not take beam tails into account.

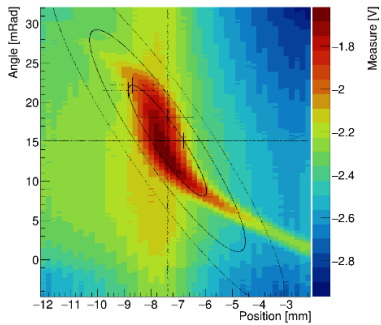

FIG. 10. Example of a emittance measurement with a non-Gaussian beam. The beam shows two tails, the first at the top, the other in the bottom.

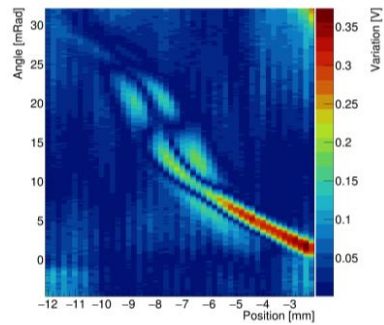

FIG. 11. Absolute deviation between fit and raw data. In this example, the fit doesn't consider the tail of the beam on the lower right part.

Data was analyzed with different beam settings (focusing with Q44 quadrupole). The new method is compared to a conventional technique using simple background filtering (pedestal), see Fig. 12. These settings have no impact on the emittance values. It can be seen in Fig. 12 that the 2D Gaussian fit gives a more consistent result with reduced fluctuations compared to the linear fitting of rms-emittance and the usual threshold variation algorithm [26]. The 2D Gaussian fit has the advantage that it is less dependent on the offset that some data may have and on the local noise fluctuations. The uncorrelated part of the background is defined during the first steps of the algorithm. In the case where the measurements contain only a background, an average Pearson chi-square of 0.74 ± 0.28 is obtained between the generated surface and the data. In comparison, in the presence of the beam, an average Pearson chi-square of 344 ± 27 reflects the accuracy of the background characterization using extreme positions and angles.

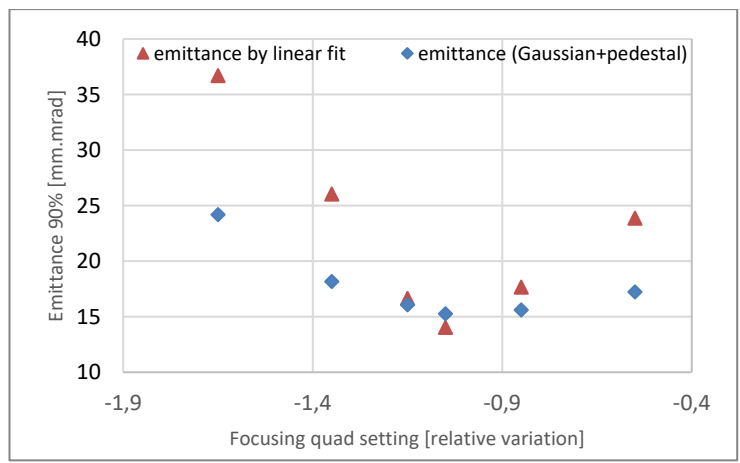

FIG. 12. Emittance variation with different beam focusing as obtained by the 2D Gaussian fit (blue dots) and the linear fit (orange dots). 2D Gaussian fit seems more stable than the linear fit (threshold variation method not represented).

It needs to be noted that a contribution at higher angles near the RIO was observed on some emittance measurements after processing the uncorrelated parts of the background, see the white ellipse Fig. 13. This low amplitude signal is not considered part of the beam because it is not present in the emittance measurements of the pepperpot unit installed upstream the Allison emittance meter, see Fig. 6. Also, this signal cannot be generated after processing the uncorrelated part of the background because it is a correlated part of the raw data. According to the literature [27], and given its position above the beam in the phase-space measurement, this component could be a "ghost" induced by the beam hitting one of the deflection plates of the emittance meter.

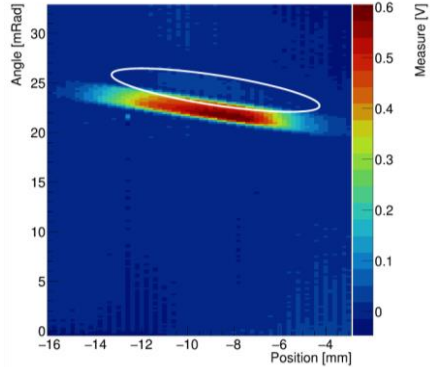

FIG. 13. Low signal detected above the main RIO and identified by the white ellipse as a "ghost".

**5.2 Data analysis for the pepperpot emittance meter**

Emittance acquisition with the pepperpot emittance-meter consists of sampling multiple bands of beams through the movable pepperpot mask along the sampling direction. The resulting datafile is a concatenation of CCD camera pictures on the form of a 2D histogram. Each independent picture is a 200x1200 pixels histogram representing the quantity of particles reaching each pixel through the 60 mm drift between the mask and the MCP. The 20 µm diameter holes from the mask should be considered as points for the data analysis.

Each camera pixel can then be linked to a position (center of the corresponding hole) and an angle with simple trigonometry (angle = Arctan (distance pixel-to-hole / distance mask-to-MCP)) in both vertical and horizontal axes. The sampling axis (vertical in our case) has a better position accuracy as the mask positions can be sampled down to 0.1 mm steps depending on the beam spot size along this direction, while the other axis has a fixed hole spacing (0.5 mm or 1 mm depending on the mask model). Angle accuracy is the same for both axes as it depends only on the CCD pixel size and drift distance but the sampling axis has a bigger angular acceptance as there can be no overlapping between the holes data.

Noise treatment is a critical issue as the absolute emittance values greatly depends on how it is handled. We used a standard method that consists of increasing progressively the analysis threshold on the data histogram until the emittance value reaches a linear regime [10]. Absolute geometrical emittance can then be extracted as shown in Fig. 14 from the extrapolation of the linear fit to a null threshold value.

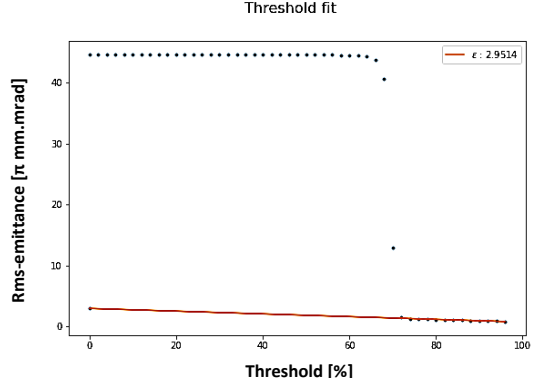

FIG. 14. Linear fit allowing the extrapolation of the emittance value for a threshold at 0% (threshold taken as a percentage of maximum beam current intensity).

## 6 Discussion

One of the main advantages of the pepperpot system is to measure the complete 4D space-phase distribution allowing the definition of the correlation coefficients between the different phase planes *(xy', yx', x'y')*, see Fig. 15. According to Liouville's theorem, 4D emittance is invariant of the motion at constant energy and the coupling between the phase-space planes in LEBT lines are usually considered negligible with good approximation and therefore 2D emittances are constant. Due to the possible beam coupling in the case on non-linear fields, use of ECRIS, solenoid but also hexapole magnet and RF kicker, some inter-plane correlation moments become nonzero and the 2D projected emittances are modified [28-35].

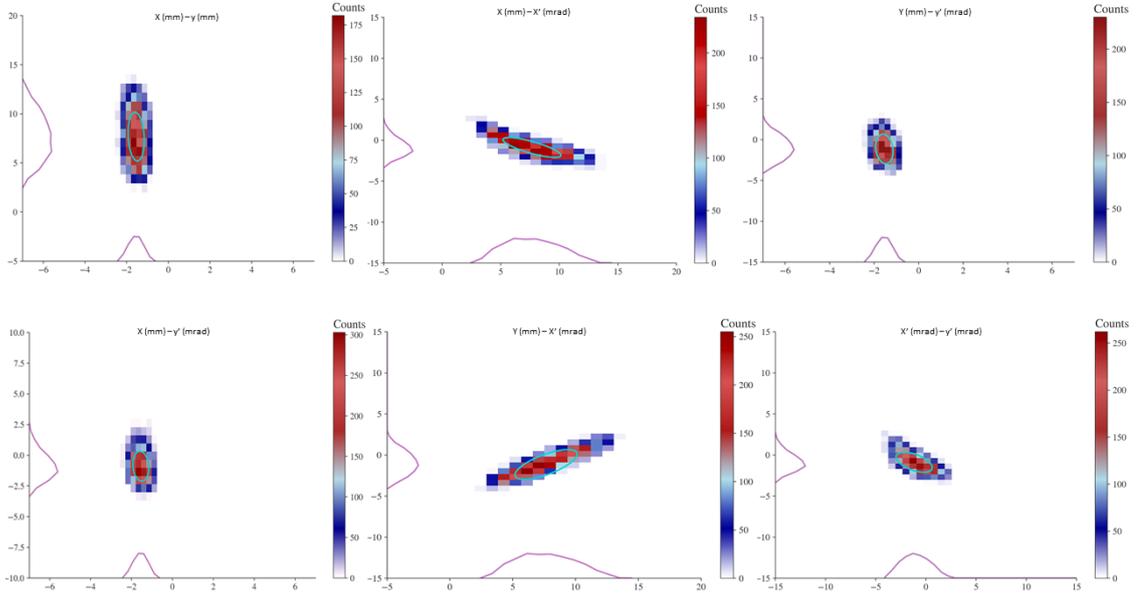

FIG. 15. Different projections of the 4D trace-space distribution in the *xy, xx', yy', xy', yx',* and *x'y'* planes. The study case corresponds to the last series shown in Fig. 8 with reduced slit opening. The first figure on the top left shows the cross section of the beam, the two next show the 2D emittances, and the three last on the bottom show the interplane correlation moments *xy', yx'* and *x'y'*. The 2D emittance in the *xx'* plane (horizontal) shows an elliptical shape with a curved axis. This is typical of axisymmetric lenses like quadrupoles exhibiting spherical aberrations (multipole fields with high order aberrations *i.e.* index $n \geq 3$). The same type of aberration is visible in the *x'y'* phase plane which reflects a coupling between these two moments.

The 6-dimensional hyper volume *V6* occupied by *N* particles in the *(x, y, z, p_x, p_y, p_z)* phase-space is defined by:

$$V_6 = \iiiint\!\!\int\!\int dx \cdot dy \cdot dz \cdot dp_x \cdot dp_y \cdot dp_z. \quad (2)$$

The conservation law and Liouville's theorem give equation (3) with the assumption of conservative forces (analogy with incompressible fluid):

$$\Delta V_6 = 0. \quad (3)$$

Equation (4) holds if there is no coupling between transverse and longitudinal motions (4). Equation (5) is applied with the assumption of no coupling between two transverse motions as for the "Allison" emittance meter but this assumption is not always consistent with a real beam:

$$V_6 = \iiiint dx \cdot dy \cdot dp_x \cdot dp_y \cdot \iint dz \cdot dp_z. \quad (4)$$

$$V_4^T = \iint dx \cdot dp_x \cdot \iint dy \cdot dp_y. \quad (5)$$

Finally, equation (6) is used to convert the phase-space to the trace-space, $p_x$ becomes *x'* and $p_y$ becomes *y'* for practical use as measurement of the gradient of the trajectories in *xz* and *yz* planes replaces transverse moments:

$$V_4^T = \iiiint dx \cdot dy \cdot dp_x \cdot dp_y = \vec{p}^2 \cdot \iiiint dx \cdot dy \cdot dx' \cdot dy' = \vec{p}^2 V_4^T. \quad (6)$$

The complete definition of the 4D emittance with all the coefficients of the beam matrix requires two "Allison" devices mounted perpendicularly on the beam line in the *xx'* and *yy'* planes supplemented by an intermediate azimuthal position in order to define the beam matrix cross coefficients (usually taken at 45°). One or two solenoids or skew quads can be used to decorrelate the affected moments of the beam matrix and thus modify the projected 2D emittances. The product of projected 2D emittances can be larger than 4D one but after solenoids and skew quads the product can be minimized in order to reach a ratio of unity between the product of the 2D emittances $\varepsilon_x$ and $\varepsilon_y$ and the 4D emittance. In this case, the interplane correlation moments will be almost completely removed. The transfer between the 2D emittances $\varepsilon_x$ and $\varepsilon_y$ shows a periodical behavior depending on the focusing strength of the solenoids or skew quads (current setting). The modification of the projected emittances with preservation of the full 6D emittance and transfer from one plane to the other is a rather recent issue. It led to the development of flat beam concept with one emittance in one plane much smaller than the other one in the other plane. It found application for electron beam in 2001 and for ECRIS in 2006 to achieve high mass resolution for RIB's [36]. The high transmission efficiency of the beam extracted from ECRIS requires a good understanding of transverse coupling.

There are many experiments with Allison emittance-meters for which coupling can be neglected or ignored, see for example studies carried out at SNS [20-22], FAIR [37], FRIB [38], TRIUMF [39], GANIL [40-41], LNL [42], ITER [43]. This doesn't mean that it is not a problem but that the coupling issue must be relativized in comparison of other sources of error which contribute to the result and can be dominant as for example ghost signal (contaminants, neutrals, electrons, etc.), noise and background, space charge and dynamic neutralization (with compensation), accuracy of backtracking simulation, and ability to measure low charge density.

Nevertheless, the measurement of only one or separately of the two transverse emittances with "Allison" emittance-meter does not allow to characterize the beam completely and to carry out effective transmission. It should be mentioned that:
  i. Accurate beam dynamics, beam matching and emittance comparison along beam line according to Liouville's theorem are not always an issue. Deleterious effect can be attenuated if light ions are considered, with symmetric beam shape and adequate beam line settings.
  ii. The solenoid used in the injection channel of the LEBT lines generates a beam coupling which contributes or compensates the coupling produced by ECRIS depending on its focusing strength and its polarity. Optimal settings are possible and solenoid polarity reversal can counteract the increase in 2D emittance and lead to minimal product of the 2D emittances $\varepsilon_x$ and $\varepsilon_y$.
  iii. Beam coupling can be eliminated by additional solenoid and skew quads (skew magnets allow 4D emittance measurement with the 2D Allison system but this is time consuming).
  iv. Coupling can be determined with 2D emittance-meters and measurements in both transverse planes if they are combined with backtracking simulations (accuracy depends on the beam line model and beam current stability).
  v. The coupling can also be determined with three emittance meters, one of which is turned 45 degrees. Another option is to use a single device mounted on a rotating flange, such as the GSI system called ROSE [44].

# 7 Summary and conclusion

We presented the transverse emittance measurements carried out with two of the most common systems on a low energy ion-beam line and performed a benchmarking. The first scanner is based on the so-called Allison principle and was developed at IPHC, and the second is a commercially available pepperpot system. The IPHC scanner has a high resolution allowing the tail of the beam to be defined but not the halo. The pepperpot system is fast, useful for beam alignment and allows definition of coupling between transverse phase planes. The difficulties in performing absolute emittance measurements and mitigating the influence of the background noise and optical aberrations in the emittance figures have been highlighted. A new technique of data filtering, noise extraction and image reconstruction is proposed in order to improve emittance figures quality and RMS calculations because standard methods like thresholding and background subtraction turned out to be of insufficient accuracy with non-Gaussian beams and noisy signals. The new method is able to define the tail of the beam (low charge density region) and detect tiny beam components (ghost signals). Finally, the interest of being able to measure the coupling between planes of the phase-space with the 4D pepperpot system and the replacement solutions with the 2D "Allison" type scanner have been discussed. We recall that 2D emittance measurements complemented by experiments and simulations can fill the lack of measurement of interplane correlation moments in case of beam coupling.

**Acknowledgements**

We acknowledge the experiment P1246-A granted by the GANIL iPAC and the ARIBE facility that are jointly run by the GANIL and CIMAP laboratories. We thank for provision of ion beams and instrumentation. We also acknowledge the 300 kV implanter of the ACACIA facility at iCUBE laboratory, CNRS/INP Strasbourg, which was used for the commissioning of the equipment.
    The data analysis was partly supported by grants from the French National Agency for Research, ARRONAX-Plus n°ANR-11-EQPX-0004, IRON n°ANR-11-LABX-18-01, and ISITE NExt no ANR-16-IDE-0007. It is supported in part by a PhD scholarship from the Institute of Nuclear and Particle Physics (IN2P3) from the National Scientific Research Center (CNRS).